\documentstyle[12pt]{article}
\input{psfig}
\begin{document}

\centerline{\large\bf Few-Body States in Lund String Fragmentation Model}
\baselineskip=22pt
%\centerline{\footnotesize\bf I. Formalism}
%\baselineskip=16pt
%\centerline{\footnotesize  Work}
%\centerline{\footnotesize E-mail:markus@thep.lu.se}

%\vfill
%\vspace*{0.6cm}
\centerline{\footnotesize\bf Bo Andersson$^1$, Haiming Hu$^{1,2}$}
\baselineskip=13pt
\centerline{\footnotesize\it $^1$Department of Theoretical Physics, 
University of Lund,
S\"olvegatan 14A, 22362 Lund, Sweden}
\centerline{\footnotesize\it $^2$ Institute of High Energy Physics,
Academia Sinica, Beijing 10039, China}
\baselineskip=12pt
%\centerline{\footnotesize\it City, State ZIP/Zone, Country}
\centerline{\footnotesize E-mail: bo@thep.lu.se }
\vspace*{0.3cm}

%\vfill
\vspace*{0.9cm}
\abstract{The well-known Monte Carlo simulation packet JETSET is not
built in order to
describe few-body states (in particular at the few GeV level in $e^+
e^-$ annihilation as in BEPC). In this note we will develop the
formalism
to use the basic Lund Model area law directly for Monte Carlo
simulations.}

%\abstract{The well-known Monte Carlo simulation packet JETSET is not
%useful to
%describe few-body states (in particular at the few GeV level in $e^+
%e^-$ annihilation as in BEPC). In this note we will develop the
%formalism
%to use the basic Lund Model area law directly for Monte Carlo
%simulations.}
 
%\vspace*{0.6cm}
\normalsize\baselineskip=15pt
\setcounter{footnote}{0}
\renewcommand{\thefootnote}{\alph{footnote}}

\section{Introduction}
The Lund Fragmentation Model contains a set of simple assumptions. Based
upon
them one obtains as a final
result an area law for the production of a set of mesons from a
string-like force field. It is possible to reformulate this into an
iterative cascade process in which one particle is produced at a time.
This is
essentially the way the model is implemented in the JETSET Monte Carlo 
packet$^{\cite{TS}}$. 

In
all computer programs it is necessary to make certain approximations,
but the
approximations in JETSET are hardly noticeable against the background
noise
signals as soon as sufficiently many particles are produced. For the
few-body
states (in particular at low energies) this is, however, no longer the
case and
in order to treat these situations it is necessary to use other means to
implement the Lund Model as a Monte Carlo simulation program. In this
note we
will show how to make use of the basic area-law directly. We will be
satisfied
to treat two-body up to six-body states, which constitutes the
overwhelming
amount of the data obtained at the BEPC/BES accelerator in Beijing.

We will start by briefly reviewing some features of the string
fragmentation scheme
in the Lund Model. After that we will in the next section show how to
implement the basic area law and present the Monte Carlo program LUARLW.
Finally we will show some results, in particular point to some places 
where there are
major deviations between the results from JETSET and LUARLW.

Although the formulas in the Lund Model are derived$^{\cite{BoSBAGG}}$,
using
(semi-)classical probabilities the final results can by comparison to
different
quantum mechanical processes be extended outside this 
framework$^{\cite{BA}}$.
We will in this paper neglect all gluonic emissions and concentrate on
the 
situation when the (color) force field from an original 
quark($q_0$)-antiquark($\bar{q}_0$) pair, produced in e.g. an
$e^+e^-$ annihilation 
event, decays into a set of final state hadrons. These are usually
termed
two-jet events. The transverse momentum of the
final state particles will be treated in accordance with a tunneling
process,
which leads to gaussian transverse fluctuations, which are 
governed by the strength of the string constant $\kappa$.
\begin{figure}[tb]
  \hbox{
     \vbox{
        \begin{center}
        \mbox{
        \psfig{figure=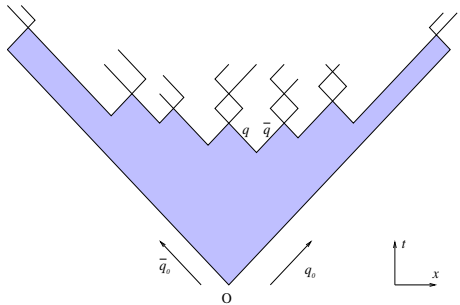,height=4cm,width=6cm,angle=-90}
        }
        \end{center}
    }
  }
  \caption{\em String fragmentation and mesons production in $t-x$ space
} 
  \label{breakup}
\end{figure}

The color force fields are in the Lund Model modeled by the massless
relativistic string with color $3$($q$) and $\bar{3}$($\bar{q}$) at the
endpoints (the gluons ($g$, color$8$) are treated as internal
excitations on the string field).
This means that there is a constant force field ($\kappa \simeq
1~GeV/fm$, corresponding to a linearly rising potential) spanned 
between the original pair. This pair is produced at the
origin $O$, cf. Fig.\ref{breakup}, and afterwards the $(q_0\bar{q}_0)$
are 
moving apart along the
$x$-axis (the longitudinal direction). 
The energy in the field can be used to
produce new $q\bar{q}$-pairs (new endpoints, thereby breaking the
string). The 
production rate of such a pair (mass(es) $\mu$, transverse momentum $\pm
{\bf k}_{\perp}$, transverse mass(es)
$\mu_{\perp}=\sqrt{\mu^2+{\bf k}_{\perp}^2}$ and with combined internal
quantum
numbers corresponding to the vacuum) is from quantum mechanical
tunneling in 
a constant force field equal to
\begin{eqnarray}
\label{tunneling}
\exp\left(-\frac{\pi \mu_{\perp}^2}{\kappa}\right).
\end{eqnarray}
The final state mesons in the Lund Model correspond to 
isolated string pieces containing 
a $q$ from one breakup point (vertex) and a $\bar{q}$ from the
adjacent vertex together with the produced transverse momentum and the
field
energy in between. 

In order to simplify the formulas and the pictures we will
treat all $q\bar{q}$-particles as massless and therefore as moving along
light-cones (massive
$q\bar{q}$-particles move in this semi-classical scenario along
hyperbolas
with these light-cones as asymptotes but the final results are the 
same$^{\cite{BA}}$).
For the longitudinal dynamics in the
Lund Model this is in accordance with quantum mechanics 
(in practice we only use dynamical features 
where position and rapidity of the particles are the relevant variables
and
this is allowed by the indeterminacy relations$^{\cite{GottLow}}$). 

One necessary requirement is that to obtain real positive (transverse) 
masses all the vertices
must have spacelike difference vectors. Together with Lorentz invariance
this means that all the vertices in the production process must be
treated in
the same way$^{\cite{BoSBAGG}}$. Another consequence is that it is 
always the slowest mesons 
that are firstly produced in any Lorentz frame (corresponding to the
fact that
time-ordering is frame dependent; this is also in accordance with the
well-known
Landau-Pomeranchuk formation time concept). Further each vertex has the
property, cf. Fig. \ref{breakup}, that it 
will divide the event into two jets; the mesons produced 
along the string field to the right and those produced 
to the left of the vertex. This feature implies that it is consistent to
introduce a rank-ordering along the light-cones so that the first rank
particle
contains the original $q_0$ (or $\bar{q}_0$) together with the $\bar{q}$
($q$)
produced at the first vertex etc. Whether the process is ordered along
one or
another of the light-cones should be irrelevant. 

Based upon these assumptions it is possible to
obtain a unique stochastical process for the decay 
$^{\cite{BoSBAGG},\cite{BA}}$. In
particular this process will turn out to have simple 
factorization properties so that for every
finite energy, rank-connected group of final state particles 
(which may possibly be part of the result from a large
or even infinitely large energy reaction) there is a factor
corresponding to
the probability to produce
the cluster with a certain energy and another corresponding to the 
probability that it decays into just
this particular final state. We will from now on concentrate on this
latter
probability distribution.

We find that the (non-normalized) probability for the cluster to decay
into the
particular channel with the $n$ particles $\{p_j\}$ is given by
\begin{eqnarray}
\label{internal}
dP_{int}= [\prod\limits_{j=1}^{n} N_j d^2p_j] \delta(p_j^2 - m_j^2) 
\delta (\sum\limits_{j=1}^{n} p _j-P_{tot})\exp(-b{\cal A}).
\end{eqnarray}
We note the appearance of the phase space for the final state particles
multiplied
by the exponential area (${\cal A}$ is the shadow region in Fig.
\ref{breakup}) decay law. 
The quantity $P_{tot}$ is the total energy
momentum of the cluster so that $P_{tot}^2=s$, $b$ is a fundamental
color-dynamical parameter and $N_j$ are normalization
constants. Actually it is possible to generalize this result and include
some
special vertex factors but just as it is in connection with JETSET 
we have found no necessity to use such generalizations for
the results in this paper.

We will start the investigation assuming that there is 
a single particle species with mass $m$. Due to Lorentz
invariance the integral of the
distribution $dP_{int}$ can only depend upon $s$  
and this means that we can define the
function $R_n$ for the multiplicity $n$ by
\begin{eqnarray}
\label{gndef}
\int dP_{int}(p_1,\cdots,p_n)=R_n(s).
\end{eqnarray}
 It is possible to subdivide these integrals into one part, that 
 will depend solely upon
the first-rank particle $p_1$ and a remainder part. 
The phase space element can be written as $dp_1
\delta(p_1^2-m_1^2)= dz_1/z_1$ with $p_{+1}=z_1P_{+ tot}$ and
$p_{-1}=m_1^2/p_{+1}$.
Similarly we find for
the area 
\begin{equation}
{\cal A}(1,\cdots n)= m_1^2/z_1 +{\cal A}(2,\cdots n).
\end{equation}
Finally the energy momentum $P_1$, obtained
after taking away the first particle, fulfills 
\begin{equation}
(P_1)^2 \equiv s_1 =
(P_{tot}-p_1)^2 = (1-z_1)(s-m_1^2/z_1).
\end{equation}
Therefore we obtain an integral
equation
linking $R_n$ to $R_{n-1}$:
\begin{eqnarray}
\label{unitarity}
R_n(s)= \int N_1 \frac{dz_1}{z_1} \exp(-bm_1^2/z_1) R_{n-1}(s_1).
\end{eqnarray}
If we define the function $R(s)= \sum\limits_n R_n(s)$ the Eq
(\ref{unitarity})
will
be valid also for $R$ itself
(to be precise there is a lower cutoff in $z$ to make it possible to
produce
more than one particle). The equation has an (asymptotic) solution $R
\propto 
s^a$ if 
\begin{eqnarray}
\label{fnorm}
1= \int N_1 \frac{dz}{z} (1-z)^a \exp(-bm_1^2)\equiv \int f(z) dz.
\end{eqnarray} 
We have neglected a
finite energy correction term $(1-m_1^2/zs)^a$ in the asymptotic limit. 
This defines the fragmentation function 
$f$ in the Lund Model and this is the way JETSET implements the model.

To be more precise, JETSET contains the production of all known
particles. 
It is assumed that the parameters $b$ and $a$ are
fixed by experiments (ordinary default values in the LEP region are
$b\simeq
0.6$ $GeV^{-2}$ and $a \simeq 0.3-0.5$) and that $N_j$ normalizes the
fragmentation function for each given mass $m_j$. The relative
probability of
the different particle flavors is decided by an external procedure. Thus
the
production of
strange quark-pairs ($s\bar{s}$) are suppressed compared to $u$ and $d$ 
(cf. Eq.(\ref{tunneling}),
an ordinary default value for the suppression factor 
is $0.3$). It is further assumed
(and this can be verified on a qualitative level) 
that it should be more difficult for a quark to tunnel into a vector 
meson (tensor meson) state (in the string model essentially a
one-dimensional  
bag) than
into a pseudoscalar state. For the production of baryons there are
several more
parameters (and recent investigations$^{\cite{PEGG}}$ indicates that for a
description of baryon production one needs to introduce special
features).

There is another possible interpretation which has been pursued by a
group from
UCLA$^{\cite{CDB}}$. In that case all the numbers $b$, $a$ and $N$ are
taken as
experimentally determinable parameters 
and the size of the fragmentation function integral in Eq.
(\ref{fnorm}) is used to determine the relative weight for the
production of
different particles. They use some straightforward Clebsch-Gordan
coefficients,
take into account that strangeness and baryon number must be conserved
(thus
e.g. the production of pairs of strange mesons must 
be used to determine the relative weight) etc. and with very few
parameters
they seem to obtain a phenomenologically good fit to data. 

There is, however, in both these implementations 
the problem of how to end the cascade process.
In JETSET (and UCLA is using a similar procedure) the process is pursued
stochastically from both endpoints and continues until there is  
a remainder mass of a certain size. This remainder is then
fragmented as a two-particle state. This ``joining''-procedure is
implemented with great care in JETSET but nevertheless provides some
unwanted
features in particular for few-body states and low energies which is the
main subject of this note. In the next section we will present a
different
procedure.

\section{Formalism}

Our basic assumption will be that the area-law, supplemented with
certain
simple probabilities similar to JETSET, will be fundamental also for
low-energy
and few-body states. We will further use a gaussian 
transverse momentum spectrum and
the procedure is outlined in Appendix 1.
 We will for definiteness
order the particles in rank
along the positive light-cone.

A particularly simple case is the two-body decay because in this case
energy-momentum conservation determines all the observables. We obtain
immediately for the quantity $R_2$ (using $m_j$, $j=1,2$, for the
transverse
masses; note that transverse momentum conservation in this case implies
that
${\bf p}_{\perp 1} = - {\bf p}_{\perp 2}$)
\begin{equation}
\label{R2body}
R_2 =N_1N_2\frac{2 
exp[-\frac{b}{2}(s+m^2_1+m_2^2)]}{\sqrt{\lambda(s,m_1^2,m_2^2)}}
\cosh(\sqrt{\lambda(s,m_1^2,m_2^2)}~),
\end{equation}
where the $\lambda$ is defined as
\begin{displaymath}
\lambda(x,y,z)= x^2 +y^2+z^2 - 2xy -2xz -2yz. 
\end{displaymath}
The two different terms correspond to the two different values that can
be
chosen for the positive light-cone fraction $z_j=p_{+j}/P_+$
\begin{eqnarray}
\label{z2body}
z_1^{\pm}&=& \frac{s+m_1^2-m_2^2 \pm \sqrt{\lambda
(s,m_1^2,m_2^2)}}{2s},
\nonumber\\
z_2^{\pm}&=& \frac{s+m_2^2-m_1^2 \mp
\sqrt{\lambda (s,m_1^2,m_2^2)}}{2s}.
\end{eqnarray}
We note that the two solutions will fulfill the relation $z_j^+ z_j^- s=
m_j^2$.
This evidently means that we obtain the corresponding negative light-cone
fractions by exchanging the sign in front of the square root.
Although the expression for $R_2$ looks singular at the  threshold 
$\sqrt{s}=m_1+m_2$, it is in practice finite and even vanishing if the
transverse momentum fluctuations are taken into account.

From the expression for $R_2$ we may derive the expression for $R_3$
according
to Eq. (\ref{unitarity}) but it is actually more convenient to consider
$R_3$
as a density in the area-size ${\cal A}_3$. Again using the positive light-cone
fractions $z_j$ ($j=1,2,3$) we may write the energy-momentum
conservation
equations and the size of ${\cal A}$ as
\begin{eqnarray}
\label{3bodyA}
\sum\limits_{j=1}^{3} z_j=1, \nonumber \\
\sum\limits_{j=1}^{3} \frac{m_j^2}{z_j}=s, \nonumber \\
\frac{m_1^2}{z_1}+\frac{m_2^2}{z_2}(1-z_1)+m_3^2={\cal A}.
\end{eqnarray}
Performing the integral we obtain
\begin{equation}
\label{R3body}
dR_3 =\prod\limits_{j=1}^{3}N_j \frac{\exp(-b{\cal A}) 
d{\cal A}}{\sqrt{\Lambda}},
\end{equation}
where,
\begin{eqnarray*}
\Lambda&=&[(s-{\cal A})({\cal A}
-m_{\perp 1}^2-m_{\perp 2}^2-m_{\perp 3}^2)\nonumber\\
 &-&m_{\perp 1}^2m_{\perp 2}^2-m_{\perp 2}^2m_{\perp 3}^2
-m_{\perp 3}^2m_{\perp 1}^2]^2 
-4sm_{\perp 1}^2m_{\perp 2}^2m_{\perp 3}^2.  
\end{eqnarray*} 
A closer investigation tells us that the area ${\cal A}$ in this
three-body
case will
be limited to the region
\begin{eqnarray}
\label{A3limits}
{\cal A}_3^- &\leq &{\cal A} ~\leq~ {\cal A}_3^+, \nonumber\\
{\cal A}_3^{\pm} &=& \frac{s+U \pm \sqrt{(s-U)^2-4(V+\sqrt{W})}}{2},
\end{eqnarray}
and that the quantity $\Lambda$ can be written as 
\begin{equation}
\label{Lambda}
\Lambda = ({\cal A}_3^+-{\cal A})({\cal A}-{\cal A}_3^-)
[(s-{\cal A})({\cal A}-U)-V+\sqrt{W}),
\end{equation}
where we have introduced the notations 
\begin{eqnarray}
\label{UVW}
U&=&m_1^2+m_2^2+m_3^2,\nonumber\\
V&=&m_1^2m_2^2+m_2^2m_3^2+m_3^2m_1^2,\nonumber\\
W&=&4sm_1^2m_2^2m_3^2.\nonumber
\end{eqnarray}
In this way the limits on the area corresponds to two of the zeros of
the
quantity $\Lambda$. It is rather easy to see that the particle
configurations
corresponding to ${\cal A}={\cal A}_3^{\pm}$ corresponds to the 
cases when the vertices
(the
breakup points) both lie on the same hyperbola.
This means in particular
that
all the possible areas for the three-body decay are in between the two
hyperbolas. To see that it is useful to
note that the square root in the definition of ${\cal A}_3^{\pm}$ can be
factorized as
\begin{eqnarray}
\label{rootfact}
(s-U)^2-4(V+\sqrt{W})&=&[\sqrt{s}-m_1-m_2-m_3][\sqrt{s}+m_1+m_2-m_3]
\nonumber\\
                   &\times&
[\sqrt{s}+m_1-m_2+m_32][\sqrt{s}-m_1+m_2+m_3]
\end{eqnarray}
and that it is a direct generalization of the corresponding expression
for the
two-body distribution
\begin{equation}
\lambda(s,m_1^2,m_2^2)=[\sqrt{s}-m_1-m_2][\sqrt{s}+m_1+m_2] 
[\sqrt{s}+m_1-m_2][\sqrt{s}-m_1+m_2].
\end{equation}
There are two sets of solutions for the light-cone fractions for given
masses,
energy and area size ${\cal A}$ (just as for the two body case). They can be
written as
\begin{equation}
\label{z1}
z_1=\frac{-B\pm \sqrt{\Lambda}}{2s(A-m_{\perp 2}^2
-m_{\perp 3}^2)},
\end{equation}
where,
\begin{eqnarray}
B&=&A(A-m_{\perp 1}^2-m_{\perp 2}^2-m_{\perp 3}^2)
-s(A+m_{\perp 1}^2-m_{\perp 2}^2-m_{\perp 3}^2) \nonumber \\
&+&m_{\perp 1}^2m_{\perp 2}^2+m_{\perp 2}^2m_{\perp 3}^2+
m_{\perp 3}^2m_{\perp 1}^2,
\end{eqnarray}
the corresponding values of $z_2$ and $z_3$ can be read
out
from Eqs. (\ref{3bodyA}).

\begin{figure}[tb]
  \hbox{
     \vbox{
        \begin{center}
        \mbox{
        \psfig{figure=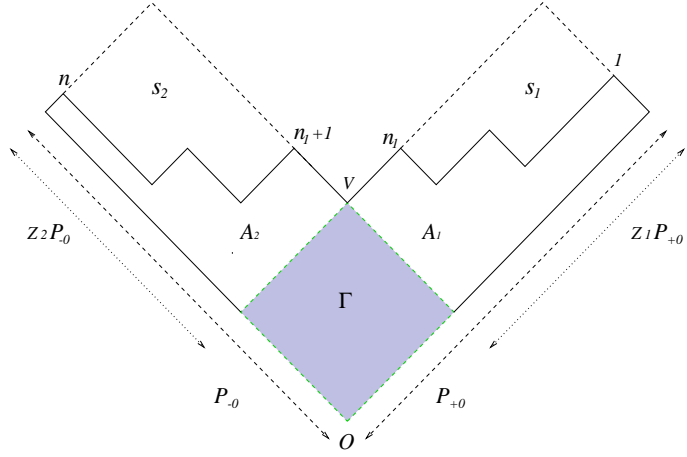,height=6cm,width=9cm,angle=-90}
        }
        \end{center}
    }
  }
  \caption{\em The vertex $V$ divides the system into two subsystems
               of ``right-movers" (mesons $1,\cdots ,n_1$) and
                  ``left-movers'' (mesons $n_1+1,\cdots ,n$).}
  \label{mlbodyarea}
\end{figure}

For the general $n$-body case we have to generate the kinematical
variables by
means of an $(n-2)$-dimensional density but as we are going to limit
ourselves
to at most six particles we will make use of a different method. We will
then
for the four-body, five-body and six-body cases subdivide the system
into two
parts, each containing $n_1$ and $n_2$ particles ($n_1,n_2=2$ or $3$ and
$n=n_1+n_2$)
. We may then, using
the simple factorization properties of the model, apply the analytical
results
obtained above for the two parts.

In more detail the subsystem closest to the positive light-cone (the
$q$-end of
the string) will be characterized by the positive light-cone fraction
$Z_+^{(1)}
\equiv Z_1$ and the other subsystem by the (negative) light-cone fraction
$Z_-^{(2)} \equiv Z_2$, cf. Fig.\ref{mlbodyarea}.
 Then the squred invariant masses of the two systems are
\begin{eqnarray}
s_1=Z_1(1-Z_2)s &{\rm and}&  s_2=Z_2(1-Z_1)s,
\end{eqnarray}
the area can similarly be subdivided into
\begin{eqnarray}
A=A_1+A_2+\Gamma & {\rm with} & \Gamma=(1-Z_1)(1-Z_2)s.
\end{eqnarray}
In this way we obtain immediately for $dR_n$ the result
\begin{equation}
\label{Rnequ}
dR_n = \frac{ds_1 ds_2}{\sqrt{\lambda (s,s_1,s_2)}}[exp(-b\Gamma_+)
+exp(-b\Gamma_-)]dR_{n_1} dR_{n_2}
\end{equation}
in terms of the two possible $\Gamma$ values
\begin{equation}
\Gamma_{\pm}=[s-s_1+s_2\pm\sqrt{\lambda (s,s_1,s_2)}~]
[s-s_2+s_1\mp\sqrt{\lambda (s,s_1,s_2)}~]/4s.
\end{equation}
The corresponding values for the $Z_j$ are (cf. the remarks after Eq.
(\ref{z2body}))
\begin{eqnarray}
Z_1^{\pm}&=&[s-s_2+s_1\pm\sqrt{\lambda (s,s_1,s_2)}~]/2s, \nonumber\\
Z_2^{\pm}&=&[s-s_1+s_2\mp\sqrt{\lambda (s,s_1,s_2)}~]/2s.
\end{eqnarray}
We note that the momentum and energies of the two subsystems as well as
their
c.m.s-rapidities are easily expressed in these variables. Consequently
the whole events can be determined in a straightforward way as soon as
the
(transverse) masses and the c.m.s energy is given. It is, however,
necessary to
introduce a method to chose the flavors of the different particles.

We will use a method ``in-between'' the JETSET and the UCLA
implementations
and in particular normalize our results at the c.m.s-energy $\sqrt{s}=4$
$GeV$
to the JETSET ones. Thus we will assume that the probability to obtain a
particular state with the $n$ mesons labeled $\{m_j\}$
(using the masses to characterize the
particles) is given by
\begin{equation}
\label{probn}
P_n (m_1,\cdots,m_n;s)=C (VPS) (SUD) N^n R_n(m_1,\cdots,m_n;s).
\end{equation}
It turns out that the functions $R_n$ are the same independent of the
rank-ordering of the particles. The quantity $C$ is then introduced as a 
combinatorial number (stemming from the quark contents of
the
possible strings-there are in general more than one such ordered
``quark-string'' possible to
make
the particular mesons and even sometimes the possibility to make the
same
meson(s) at different places in the same string). 
The quantities $(VPS)$ are the vector to pseudoscalar rate
(we neglect the tensor mesons, although they can be introduced
by straightforward means). The parameter
$(SUD)$ is the strange to up and down $(q\bar{q})$ probability. 
It comes without saying that both $(VPS)$
and $(SUD)$ depends upon the number of choices to be made. The
possibility to
produce charm in the fragmentation process is-just as in JETSET-
neglected in accordance with the results from Eq. (\ref{tunneling}).
Finally $N$ is a parameter to obtain
the right multiplicity distribution. To obtain the right 
multiplicity distribution we will  sum over all the different 
orderings and strings in Eq.
(\ref{probn}). In this way we find a default value for the parameter $N$
by comparison to JETSET.

The procedure to produce a state is then that we firstly chose the
multiplicity 
from the sum over all the $R_n$
and then the particular state by a stochastical choice
among the possible $n$-particle channels. After that we 
chose the particular ordering of
the particles stochastically from the ways they can be produced from the
possible quark-strings. Finally to obtain the particle distributions 
we just differentiate $R_n$ in the ways we have described it above.

In this way we have defined a simple few-body low-energy version of the
Lund
Model (although an incorporation of baryons into the scheme may have
some
influence upon the results). It is implemented in the Monte Carlo
simulation
program LUARLW$^{\cite{HaimingHu}}$. A list of the default parameters of
the model
is given in Appendix 2.

\section{Some Results of the Model}

We will in this section exhibit a few results from the model implemented
in
the simulation program LUARLW, in particular discuss some configurations
where
the use of JETSET may be misleading in the few-body, low-energy cases.

\begin{figure}[tb]
  \hbox{
     \vbox{
        \begin{center}
        \mbox{
        \psfig{figure=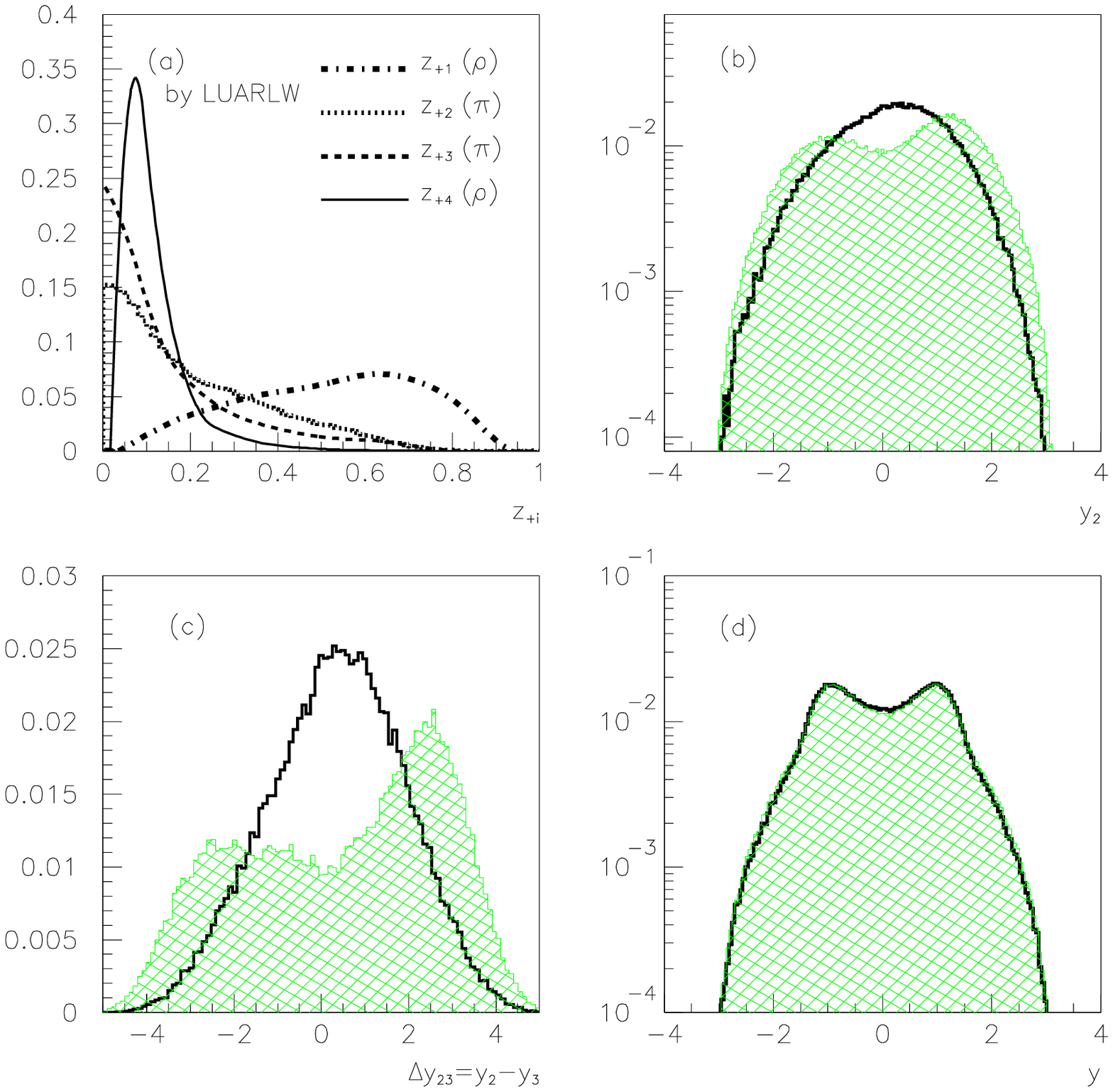,height=16cm,width=16cm,angle=0}
        }
        \end{center}
    }
  }
  \caption{\em The light-cone component and rapidity distributions 
  of a $\rho\pi\pi\rho$ final state at $\sqrt{s} = 4~ GeV$ (the ordering
  corresponds to rank-ordering).
  The distributions for the corresponding rank
  particle in JETSET (the net-like regions) and LUARLW (the solid
  lines) 
  show significant differences, while
  the total inclusive distributions agree. The rapidity-difference in
JETSET 
  shows the gap near $\Delta y_{ij}\sim 0$ (the so-called 
  ``disease'' distribution). Figure (a) is non-normalized.}
  \label{figure3}
\end{figure}

\begin{figure}[tb]
  \hbox{
     \vbox{
        \begin{center}
        \mbox{
        \psfig{figure=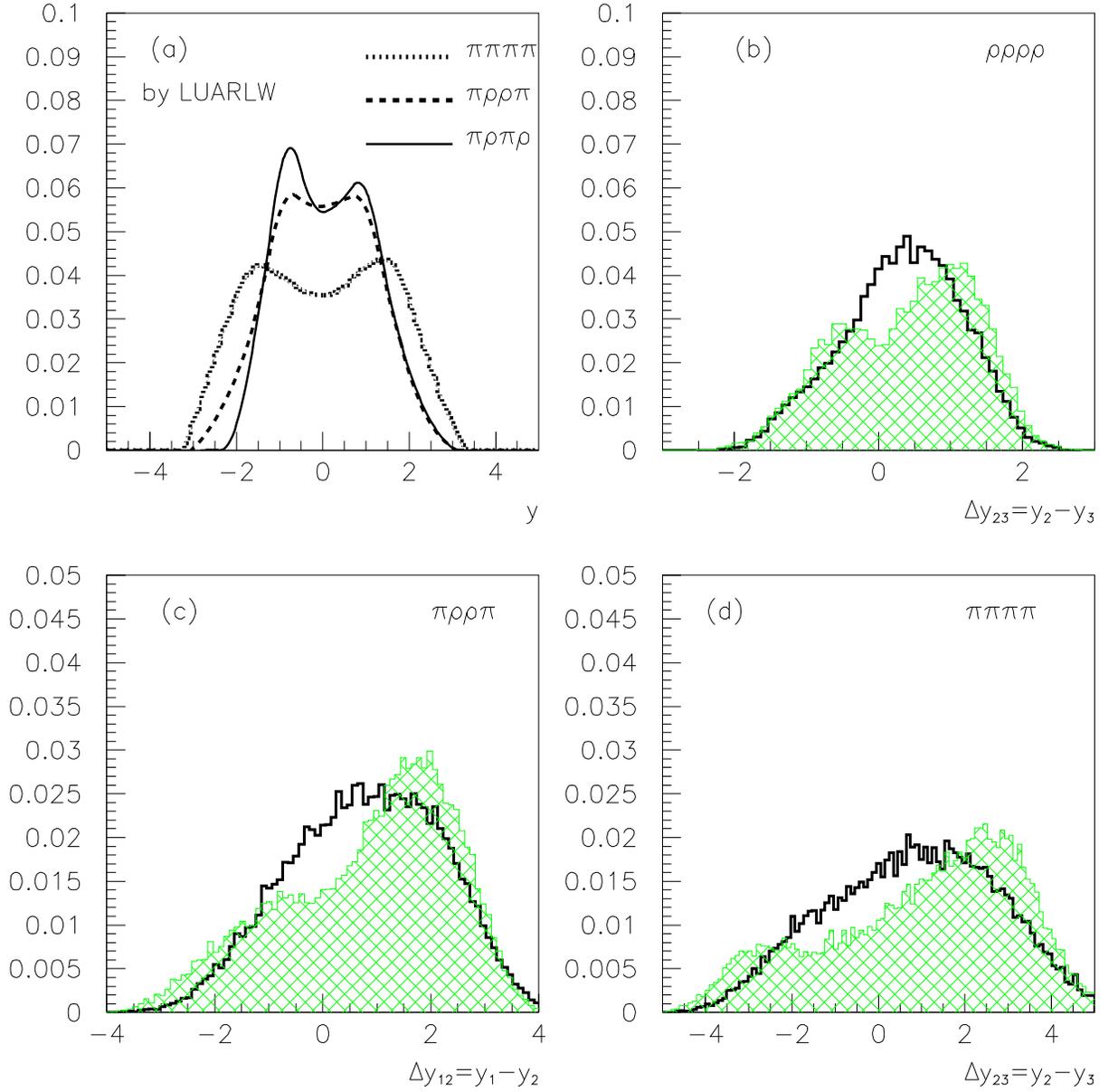,height=16cm,width=16cm,angle=0}
        }
        \end{center}
    }
  }
  \caption{\em The distributions of rapidity and rapidity
  differences for some special final states at $\sqrt{s} = 4~ GeV$ predicted
  by JETSET (the net-like regions) and LUARLW (the solid lines).}
  \label{figure4}
\end{figure}

We firstly note that in a model of iterative character it is in general
difficult to obtain the right behavior for small relative rapidities
between
two (rank-connected) particles. To see that 
we consider such a two-particle state,
that is
a part of a larger system,
and assume that the pair has a total (transverse) mass-square
$s_{12}= m_1^2+m_2^2 + 2m_1m_2\cosh(\delta y)$ with $m_j$ the individual
(transverse) masses and $\delta y$ the rapidity difference. This means
that
the distribution in $\delta y$ is
\begin{eqnarray}
\label{deltay}
\frac{dP}{d\delta y} = \frac{dP}{ds_{12}} 2m_1m_2 \sinh(\delta y)
\end{eqnarray}
Therefore unless the distribution in $s_{12}$ behaves (for values of
$\sqrt{s_{12}}$ close to $m_1+m_2$, i.e. for small $\delta y$) as
$1/\sqrt{\lambda(s_{12},m_1^2,m_2^2)}=1/2m_1m_2 \sinh(\delta y)$ then
$dP/d\delta y$ will vanish. From the results discussed above this is in
general taken into account by the generation mechanism in LUARLW but it
is not
so if we consider the corresponding situation in JETSET (cf. Fig.
\ref{figure3} and Fig.\ref{figure4}). On the other hand it is difficult
to 
observe this
effect in a
multi-particle environment so it is only for the few-body case it is of
interest, especially if the phase space is small (low energies).

There is further the ``joining mechanism'' in JETSET and related
simulation programs. One place where this is noticeable is if we consider
the
rapidity ordering of a particular set of particles as they are generated
in
LUARLW and in JETSET. As examples we have chosen the four-body states
$\rho^+\rho^-\pi^+\pi^-$ and $K^+K^-\pi^+\pi^-$.
In Table 1 we present the
probabilities for the different possible orderings in two systems.  

We have only pinpointed a few particular places where it may be possible
to see
the precise workings of the area-law as compared to an iterative
production
mechanism. We find several places where there are clear differences 
but our major finding is that it quite surprising 
how well the basic JETSET scenarium really
works at such low energies and multiplicities.

\begin{center}
\begin{tabular}{|c|c|c|c|}\hline
final state    & $\rho^+\rho^-\pi^+\pi^-$ & $K^+K^-\pi^+\pi^-$ \\\hline
rapidity order & JETSET \vline LUARLW & JETSET \vline LUARLW \\\hline\hline
$y_1>y_2>y_3>y_4$ & 13.15~ \vline ~19.78 & 18.55~ \vline ~23.10 \\\hline
$y_1>y_2>y_4>y_3$ & 12.85~ \vline ~12.39 & 14.76~ \vline ~12.85 \\\hline
$y_2>y_1>y_3>y_4$ & 12.60~ \vline ~11.94 & 14.70~ \vline ~12.42 \\\hline
$y_2>y_1>y_4>y_3$ & 12.39~ \vline ~~8.79& ~~8.47~ \vline ~~~8.86 \\\hline
$y_1>y_3>y_2>y_4$ & ~11.46~ \vline ~~12.43& 12.90~ \vline ~12.82 \\\hline
$y_1>y_3>y_4>y_2$ & ~~5.98~ \vline ~~~5.89& ~~3.72~  \vline ~~~4.98 \\\hline
$y_3>y_1>y_2>y_4$ & ~~5.96~ \vline ~~~5.96& ~~3.77~  \vline ~~~4.83 \\\hline
$y_2>y_3>y_1>y_4$ & ~~5.29~ \vline ~~~4.72& ~~6.67~  \vline ~~~4.46 \\\hline
$y_1>y_4>y_2>y_3$ & ~~5.27~ \vline ~~~5.11& ~~6.61~  \vline ~~~5.33 \\\hline
$y_3>y_1>y_4>y_2$ & ~~3.48~ \vline ~~~1.49& ~~0.41~  \vline ~~~0.09 \\\hline
$\cdots\cdots$    & ~~\vline ~  &  ~\vline  \\\hline   
\end{tabular}
\begin{table}
\caption{The probabilities ($\%$) for the different possible
         rapidity order in JETSET and LUARLW for 
         $\rho^+\rho^-\pi^+\pi^-$ and $K^+K^-\pi^+\pi^-$ final
         states.}
\end{table} 
\end{center}

\vskip1cm
\appendix
\begin{large}
\noindent
{\bf Appendix 1}\\
{\bf The transverse momentum of the final state hadrons}
\end{large}

In the Lund model, a quantum mechanical tunneling
effect has been used for the generation of
quark-antiquark pairs $q_i\bar{q}_i$. 
In JETSET,
the ${\bf p_{\perp}}$ of a meson $q_{i-1}\bar{q}_i$
is given by the vector sum of the transverse momenta of
the $q_{i-1}$ and $\bar{q_i}$ constituents.
 
In our scheme we produce the distributions of the particles directly
from the
area-law and then
the transverse momenta
$\{{\bf p}_{\perp j}\}$ of the $n$ particles must be determined at the
same
time as the longitudinal momenta $\{p_{L j}\}$. It is necessary to
conserve the
total transverse momentum and it is then necessary to include that in
the
generation. We have used two different methods which are described
below.

\section{Scheme I}

In this scheme we generate the transverse momentum of the particles
directly,
although one at the time. It is necessary to keep to total transverse
momentum conservation and then it is necessary to modify the gaussian
distribution. To that end we define the distribution
\begin{equation}
F^{(n)}({\bf p}_{\perp 1},\cdots ,
{\bf p}_{\perp n})
=\delta (\sum\limits_{j=1}^{n}{\bf p}_{\perp j})
\theta \left(\sqrt{s}-\sum\limits_{j=1}^{n}
\sqrt{m_j^2+{\bf p}_{\perp j}^2}\right)
\prod\limits_{j=1}^{n}exp(-\frac{{\bf p}_{\perp j}^2}{2\sigma^2}),\\
\end{equation}
where ${\bf p}_{\perp j}$ is a two-dimensional vector, and
$\sigma$ is the variance of the distribution.

To obtain the distribution in ${\bf p}_{\perp 1}$, we must integrate the
distribution $F^{(n)}$ over all the other vectors. We need
to
calculate the following integrals when we use the 
condition-density scheme to sample $\{{\bf p}_{\perp j}\}$.
The distribution of ${\bf p}_{\perp 1}$ regardless of
values of other transverse momentums reads,
\begin{eqnarray}
&&f_1^{(n)}({\bf p}_{\perp 1})
=\int d^2{\bf p}_{\perp 2}\cdots
d^2{\bf p}_{\perp n}F^{(n)}({\bf p}_{\perp 1},\cdots ,
{\bf p}_{\perp n})\\
&=&N_1\cdot exp[-\frac{n}{n-1}
\frac{{\bf p}_{\perp 1}^2}{2\sigma^2}].
\end{eqnarray}
Under the condition of ${\bf p}_{\perp 1}$ is fixed, the
distribution for ${\bf p}_{\perp 2}$ is
\begin{eqnarray}
&&f_2^{(n)}({\bf p}_{\perp 2}|{\bf p}_{\perp 1})
=\int d^2{\bf p}_{\perp 3}\cdots
d^2{\bf p}_{\perp n}F^{(n)}({\bf p}_{\perp 1},\cdots ,
{\bf p}_{\perp n})\\
&=&N_2\cdot exp[-\frac{n-1}{n-2}
\frac{({\bf p}_{\perp 2}+{\bf p}_{\perp 1}/(n-1))^2}{2\sigma^2}].
\end{eqnarray}
Similarly if all the vectors up to the $j-1th$ have been determined then
the
distribution  for the $jth$ $(j\le n-1)$ particle is
\begin{eqnarray}
& &f_j^{(n)}({\bf p}_{\perp j}|{\bf p}_{\perp 1},\cdots ,{\bf p}_{\perp
j-1})
=\int d^2{\bf p}_{\perp j+1}\cdots
d{\bf p}_{\perp n}F^{(n)}({\bf p}_{\perp 1},\cdots ,
{\bf p}_{\perp n})\nonumber\\
&=&N_j\cdot exp[-\frac{(n-j+1)}{(n-j)}\cdot
\frac{({\bf p}_{\perp j}+
\sum\limits_{i=1}^{j-1}{\bf p}_{\perp j}/(n-j+1))^2}{2\sigma^2}].
\end{eqnarray}
The final vector ${\bf p}_{\perp n}$ is evidently determined by
energy-momentum
conservation.
The effective variance of transverse momentum for $jth$ particle is
\begin{equation}
\sigma_{j}^{(eff)}=\sqrt{\frac{n-j}{n-j+1}}\sigma,~~~(j=1,\cdots ,n-1).
\end{equation}
We see that $\sigma_{j-1}^{(eff)}>\sigma_{j}^{(eff)}$, so the variances
will be different for the particle with different order-number.
This is because we have aleady determined a set of 
$\{{\bf p}_{\perp}\}$. If we instead ask for the inclusive
distribution in ${\bf p}_{\perp}$ (i.e. the gaussian width independent
of the rest of the particles), then it is the same for all values of $j$

\section{Scheme II}

The method in scheme I is not the most 
common approach.
In this case the transverse momentum conservation is fulfilled by
construction
and the particles obtain their transverse momenta from the constituents.
At
each production point the $(q \bar{q})$-pair is given $\pm{\bf q}$ and
the
particle momenta are then 
\begin{equation}
{\bf p}_{\perp 1}={\bf q}_1,\cdots ,
{\bf p}_{\perp j}={\bf q}_j-{\bf q}_{j-1},\cdots ,
{\bf p}_{\perp n}=-{\bf q}_n.
\end{equation} 
There is a possibility (and at least for small mass-particles like
pions this
seems to be experimentally the case) that there is a correlation in the
generation of the transverse momentums between adjacent production
points,
The most general way to introduce such a correlation in a
forward-backward
symmetric shape is, \cite{JimS}
\begin{equation}
f^{(n)}({\bf q}_1,\cdots ,{\bf q}_n)
=\prod\limits_{j=1}^{n}d^2{\bf q}_jexp-\frac{1}{2\sigma^2}
\left[{\bf q}_1^2+\frac{({\bf q}_2-\rho_2{\bf q}_1)^2}{1-\rho_2^2}
+\cdots +\frac{({\bf q}_n-\rho_n{\bf q}_{n-1})^2}{1-\rho_n^2}\right].
\end{equation}
To see the symmetry we note that it can just as well be written in the
following shape
\begin{equation}
f^{(n)}({\bf q}_1,\cdots ,{\bf q}_n)=
\prod\limits_{j=1}^{n}d^2{\bf q}_jexp-\frac{1}{4\sigma^2}\left[{\bf
q}_1^2 +
{\bf q}_n^2 +\sum A_j({\bf q}_j^2 
+{\bf q}_{j-1}^2 -2 \epsilon_j{\bf q}_j\cdot{\bf q}_{j-1})\right] 
\end{equation}
with
\begin{eqnarray}
A_j= \frac{(1+\rho_j^2)}{(1-\rho_j^2)} \nonumber\\
\epsilon_j= \frac{2 \rho_j}{(1+\rho_j^2)}
\end{eqnarray}
In the original paper \cite{JimS}, the correlation was
phenomenologically
taken as a function of the mass of the particle. In general $\rho_j$ and
therefore also $\epsilon_j$ are small numbers and can be neglected as
they are
in JETSET default.

\vskip1cm
\appendix
\begin{large}
\noindent
{\bf Appendix 2}\\
{\bf The parameters used in LUARLW}
\end{large}

In our notations, $|M|^2=exp(-bA_n)$ is understood as the
squared matrix element.
The exclusive probabilites of $n$-body final state are given by
\begin{equation}
\label{e4a1}
P_n=\bar{\sum\limits_{f}}\int d\Omega |M|^2,
\end{equation}
the following factors are considered in calculation.\\
$\bullet$~$VPS$:
A particle with spin $J$ has $2J+1$ spin projections,
which means that
the ratio of the vector mesons to pseudoscalar mesons
should be $3:1$. 
But $e^+e^-$ annihilation experiments show
this ratio is smaller than the predicted value. 
There is a dynamical reason for the vector meson
suppression$^{\cite{VPS}}$. Here, we treat the product
of $(2J+1)$ and vector meson suppression factor as free parameters,
and they may take different values for $\rho/\pi$, $K^{*}/K$ and
$D^{*}/D$ ratio.\\
$\bullet$~$C$:
In the quark model of hadron,
a specified final state may produce in the
fragmentation of several different strings. We have counted
the corresponding numbers $C$ for all channels.\\
$\bullet$~$SUD$:
Lund model invokes the idea of quantum mechanical tunnelling
to generate the $q_i\bar{q}_i$, which gives the relative probability for
$u\bar{u}:d\bar{d}:s\bar{s}\approx
1:1:r_s$. We take $r_s\simeq 0.3$ same as in JETSET.\\
$\bullet$~$N$: As a default value, $N$ is determined
by requesting LUARLW to give out
the same multiplicity distributions as in JETSET, 
\begin{equation}
N=\sqrt[n]{\frac{\sum\limits_{n fixed}P_n^{JETSET}}{
\sum\limits_{n fixed}C(VPS)(SUD)R_n}}.
\end{equation}
~~\\
\begin{center}
Table 2 The values of parameters used in Eq.(\ref{probn}) which
are fixed\\
at $\sqrt{s}= 4~GeV$ by comparison to JETSET.\\
~~\\
\begin{tabular}{|c|c|c|c|}\hline
b & VPS & SUD & N \\\hline
  & $\rho/\pi$~~ \vline ~~$K^{*}/K$~~ \vline ~~$D^{*}/D$ &  & 
    $n=3$~ \vline ~~$n=4$~ \vline ~~$n=5$~ \vline ~~$n=6$ \\\hline
0.58 & ~1.07~~~\vline~~~~1.15~~~ \vline~~~~~~-~~~~~& 0.33 & 
0.49~ \vline ~~~0.38~~ \vline ~~~0.26~~ \vline ~~0.19 \\\hline 
\end{tabular}
%\begin{table}
%\caption{}
%\end{table}
\end{center}


\begin{thebibliography}{99}
\bibitem{HaimingHu}It is implemented in the Monte Carlo simulation
                   program LUARLW, that is available on request from 
                   Haiming Hu.
\end{thebibliography}
\end{document}